\documentstyle[editedvolume, psfig]{crckapb} 

\begin{opening}
\title{ HST DETECTIONS OF MASSIVE BLACK HOLES IN THE CENTERS OF GALAXIES}

\author{H.C. Ford}
\author{Z.I. Tsvetanov}
\institute{The Johns Hopkins University\\
           Homewood Campus, Baltimore, MD 21218}
\author{L. Ferrarese}
\institute{The California Institute of Technology\\
	Dept. of Astronomy, 105-24, Pasadena, CA 91125}
\author{W. Jaffe}
\institute{Leiden Observatory\\Postbus 9513, 2300 RA Leiden, 
	The Netherlands} 

\end{opening}

\runningtitle{HST DETECTIONS OF MASSIVE BLACK HOLES}

\begin{document}

\begin{abstract}

After correcting spherical aberration in the Hubble Space Telescope in
1993, the central masses of galaxies can be measured with a resolution
5 to 10 times better than can be achieved at the best terrestrial
sites.  This improvement in resolution is decisive for detecting the
gravitational signature of massive black holes in galaxy nuclei. The
discovery of small ($r \sim 100 - 200$ pc) rotating gaseous and
stellar disks in the centers of many early-type galaxies provides a
new and efficient means for measuring the central potentials of
galaxies.  Concomitantly, VLBI observations of H$_2$O masers in the
nuclei of NGC 4258 and NGC 1068 revealed exquisite Keplerian rotation
curves around massive black holes at radii as small as 0.1 pc.  Recent
terrestrial $K$-band measurements of the proper motions of stars in
the cluster at the center of the galaxy provide irrefutable evidence
for a black hole with a mass of $2.7 \times 10^6 M_{\odot}$.  At the
time of this symposium, the presence of central massive black holes
has been established in 12 galaxies.  The evidence suggests that there
are massive black holes in the centers of all AGNs and in most, if not
all, nucleated galaxies.  The present data show at best a weak
correlation between black hole mass and bulge luminosity.

\end{abstract}

\section{HST Resolution of a Black Hole's Sphere of Influence}

HST has a stable, diffraction limited point spread function with a
full width at half maximum of $0.048''$ at 486 nm.  This resolution is
5 to 10 times better than the resolution achieved at the best
terrestrial sites.  To illustrate the importance of high spatial
resolution, consider the line of sight velocity dispersion
$\sigma_{\rm MBH}$ of stars near a massive black hole (MBH).  This
dispersion will be

		\begin{equation}				
		{\sigma_{\rm MBH}}^2 \sim GM_{\rm BH}/R.  
		\end{equation}

If the dispersion of stars in the bulge of the parent galaxy is
$\sigma_0$, the black hole's sphere of influence will be

		\begin{equation}				
		R_{\rm BH} \sim GM_{\rm BH} /{\sigma_0}^2 = 
		43~M_8/(\sigma_{100})^2~{\rm pc}
		\end{equation}

\noindent where $\sigma_{100} = \sigma_0/100$ km s$^{-1}$, and 
$M_8 = M_{\rm BH}/10^8 M_{\odot}$.

A $10^8 M_{\odot}$ MBH in a galaxy like M31 with $\sigma_0 \sim
160~{\rm km s}^{-1}$ will have $R_{\rm BH} \sim 5$~pc.  At the
distance of the Virgo cluster, $R_{\rm BH} \sim 0.2''$. Measuring the
rise in the velocity dispersion toward the center is difficult but
possible with HST, and presently impossible from the ground.

\subsection{Nuclear Disks}

In addition to providing spatial resolution unobtainable from the
ground, unexpected discoveries made with HST have revealed new ways to
measure the gravitational potential in the centers of galaxies.  HST
observations of early-type galaxies have shown that small disks ($r
\sim 100-200$ pc) of gas or stars reside in the nuclei of many active
galaxies (e.g., Jaffe et al.\ 1994 [J94], 1996). Because the disks are
much larger than the parsec scale, hot, inner accretion disks, we
refer to them as nuclear disks to avoid confusion.

Examples of the gaseous and stellar nuclear disks found in elliptical
galaxies are shown in Figure 1. Table 1 summarizes our estimates of
the disk masses derived from the H$\alpha$ luminosity and/or
extinction maps and the assumption that the disks have the same
dust-to-gas ratio as the galaxy.

\begin{quote}
\begin{table}[htb]
\begin{center}
\caption{Sizes and Estimated Masses of Gaseous Nuclear Disks}
\begin{tabular}{lllllll}
\hline
Galaxy& Distance& Radius& Radius& Inclination& $\tau_{V}$ &$M_{\rm HI+HII}$ \\
	 & (Mpc)& ($''$)& (pc)      & (Deg) &       & (M$_{\odot}$)   \\ 
\hline
NGC 4261 &  32  & 0.86  & $\sim130$ & 64    &$\sim1$   & $\sim 5\times10^4$ \\
NGC 6251 & 105  & 0.78  & $\sim400$ & 76    &$\sim1.8$ & $\sim 3\times10^6$ \\
M87	 &  15  &$\sim1$& $\sim70$  & $\sim40$&$-$     & $\sim 10^4$ \\
\hline
\end{tabular}
\end{center}
\end{table}
\end{quote}

\begin{figure}[t]
\vbox to 11cm{\vspace*{-2.75cm} \hspace*{-0.5cm}
\psfig{figure=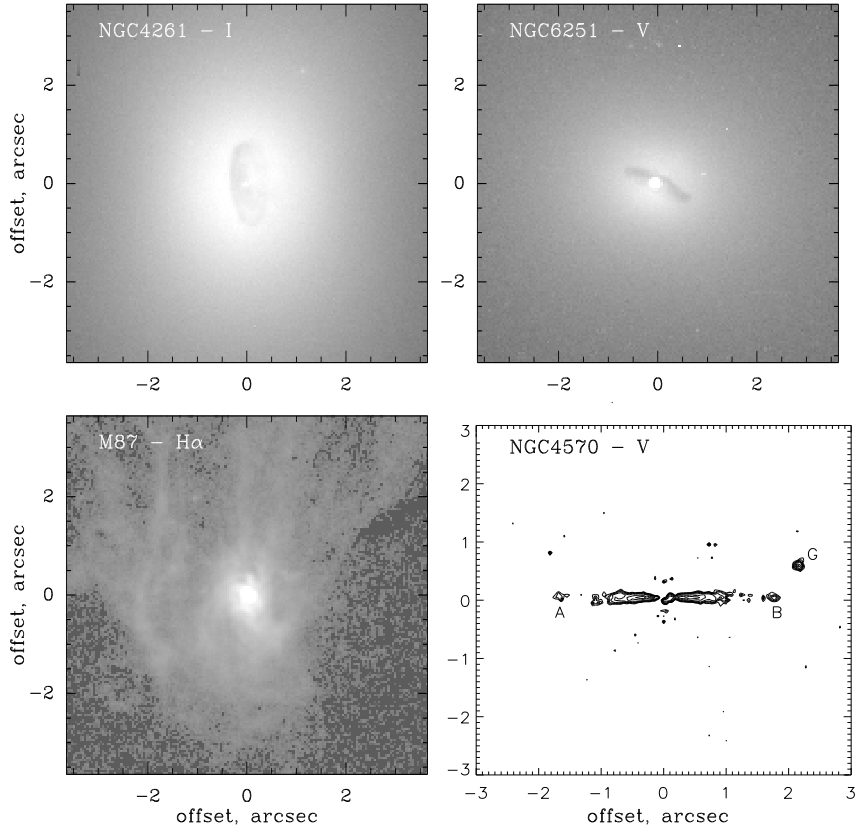,width=13cm}
}
\caption{Examples of nuclear disks in early type galaxies. Two dusty
nuclear disks are shown in the upper panels -- NGC~4261 ({\it left})
and NGC~6251 ({\it right}). The two lower panels show the H$\alpha$
disk in M87 ({\it left}) and the stellar disk in NGC~4570 ({\it
right}) after subtraction of the host galaxy.}
\end{figure}

The flattened, approximately circular nuclear disks are likely to be
in near circular rotation in the central potential of the galaxy.  The
approximate alignment between the minor axes of the disk in NGC 4261
(${\Delta}\theta = 17^{\circ}$) and NGC 6251 (${\Delta}\theta =
25^{\circ}$) and the large scale radio axis suggest a causal
connection between the angular momentum in the nuclear disk and
angular momentum in the accretion disk.  If the disk's circular motion
persists inside of the black hole's sphere of influence, measuring the
black hole's mass using emission lines in the disk will be much easier
than using the stellar velocity dispersion. This was demonstrated by
Ford et al.\ (1994) who used the newly installed WFPC2 to find a small
($r \sim 1''$;~70 pc at 15 Mpc) disk of ionized gas in the nucleus of
M87, and Harms et al.\ (1994) used COSTAR with the Faint Object
Spectrograph with a 0.26$''$ aperture to measure blue and redshifted
rotational velocities of $500$ km~s$^{-1}$ at radii $r = 18$ pc paired
across the nucleus along the apparent major axis of the disk.  These
measurements yielded a central mass of $2.4 \pm 0.7 \times 10^9
M_{\odot}$ and a mass-to-light ratio $(M/L)_V \sim 500$.  Subsequent
measurements with the FOS $0.086''$ aperture (Ford et al.\ 1996) and
the FOC (Macchetto et al.\ 1997) found Keplerian rotation around a
$\sim 3 \times 10^9 M_{\odot}$.  Likewise, FOS observations of the
nuclear disk in NGC 4261 revealed a central mass of $4.9 \pm 1.0
\times 10^8 M_{\odot}$ and $(M/L)_V \sim 2100$ within the inner 14.5
pc (Ferrarese, Ford \& Jaffe 1996).

Ford et al.\ (1997) reviewed gaseous nuclear disks in elliptical
galaxies.  They concluded that small, mostly dusty, nuclear disks and
disk like structures are common in early type galaxies, with a
frequency $\sim 30\%$.

J94 and van den Bosch et al.\ (1994) found small {\it stellar} nuclear
disks in three Type II Virgo E$\/$S0 (NGC 4342, NGC 4570, and NGC
4623).  These disks have radii $r < 100$ pc and scale lengths $\le 20$
pc.  The lower right panel in Figure 1 shows the isophotal map of the
stellar disk in NGC 4570 after subtraction of the galaxy's bulge.  As
demonstrated by van den Bosch (1997), measuring these disks' rotation
provides a means to measure the central mass.

\section{Summary of HST Observations of Massive Black Holes}

Table 2 summarizes HST estimates for dark masses in the centers of
galaxies at the time of this writing. In addition to HST measurements,
we have included the VLBA mass determinations because these
observations have approximately 100 times higher angular resolution
than achievable with HST.  We also have included the mass
determinations for the center of our Galaxy derived from ground-based
infrared spectra and proper motions of the central star cluster
(Eckhart \& Genzel 1997; Ghez et al.\ 1998; Genzel 1998).  These
observations have 20 times higher spatial resolution than can be
achieved with HST on M31 and M32.

\begin{quote}
\begin{table}[htb]
\begin{center}
\caption{High Spatial Resolution Observations of Massive Black Holes}
\begin{tabular}{llll}
\hline
Galaxy	  & Instrument 	& Resolution (mas)	& References	\\
\hline	    		  		  
NGC 6251  & FOS 	& 86	         & FFJ97 \\
NGC 4486  & FOS \& FOC	& 260, 86, 63    & H94; F96; M97 \\
NGC 4261  & FOS	        & 86	         & FFJ96 \\
NGC 4594  & SIS \&FOS	& 630(FWHM), 210 & K96a \\
NGC 4374  & STIS	& 200	         & B97 \\
NGC 1068  & VLBI	& $<$0.5         & GG97 \\
NGC 224   & FOS	        & 260, 86	 & P98 \\
NGC 3115  & SIS \&FOS	& 570(FWHM), 210 & K96b \\
NGC 4258  & VLBA	& $0.6\times0.3$ & M95 \\
Milky Way & NTT and Keck & $130-150; 50$      & EG97, Gh98 \\
NGC 4342  & ISIS \& FOS	& 1000(FWHM), 260& vdB97 \\
NGC 221   & FOS	        & 210,86	 & vdM97 \\
\hline
\end{tabular}
\end{center}
\end{table}
\end{quote}

The radial velocities of the masing disk in NGC 4258 show the
unmistakable signature of Keplerian rotation around a mass of $3.6
\times 10^7 M_{\odot}$ interior to the inner edge of the disk at a
radius of 0.13 pc (Miyoshi et al.\ 1995).  The IR observations of the
galactic center reported in these proceedings by Genzel and by Eckart
\& Genzel 1998, and Ghez et al.\ 1998, respectively find $2.7 \pm 0.2
\times 10^6 M_{\odot}$ and $2.6 \times 10^6 M_{\odot}$ inside 0.01 pc!
Maoz (1998) finds that the absolute maximum possible lifetimes of
central dark clusters in the Milky Way and NGC 4258 are respectively
$\sim 10^8$ and $\sim 2 \times 10^8$ years.  Because this lifetime is
much less than the ages of galaxies, a dark cluster is a highly
improbable source for the mass. The galactic center and NGC 4258 are
very strong cases that the central dark masses are massive black
holes.

Table 3 summarizes the central masses, assumed to reside in massive
black holes, derived from HST, VLBA, and IR observations. For each
galaxy the table lists the total blue luminosity (BT$_0$), the ratio
of the bulge/total luminosity, the bulge luminosity $L_B$, and the
assumed distance. The last column is the reference for the mass
determination method --- GD = gaseous disk, SD = stellar dynamics.

\begin{quote}
\begin{table}[htb]
\begin{center}
\caption{High Spatial Resolution Measurements of Black Hole Masses}
\begin{tabular}{lllllllll}
\hline
Galaxy	 & Type  & Dist. & BT$_0$& Bulge/&$M_{\rm Bulge}$&$M_{\rm BH}$&Method \\
	 & RSA   & Mpc   & RC3 & Total & 	  & $10^8M_{\odot}$&  \\
\hline	    	      	      	      	        	    
NGC 6251 & (E0)    & 98      &	13.22 & 1.0   & $-$21.74 & 7.50  & GD  \\
NGC 4486 & E0	   & 16      &	9.49  & 1.0   & $-$21.53 & 26.12 & GD  \\
NGC 4261 & E3	   & 32      &	11.36 & 1.0   & $-$21.17 & 4.27  & GD  \\
NGC 4594 & Sa+/Sb- & 8.9     &	8.38  & 0.93  & $-$21.13 & 9.67  & SD  \\
NGC 4374 & E1	   & 16      &	10.01 & 1.0   & $-$21.01 & 3.00  & GD  \\
NGC 1068 & Sb(rs)II& 16.2    &	9.47  & 0.23  & $-$19.98 & 0.16  & H$_2$O Masers \\
NGC 224  & SbI-II  & 0.72    &	3.28  & 0.24  & $-$19.46 & 0.75  & SD \\
NGC 3115 & S01(7)  & 6.5     &	9.74  & 0.94  & $-$19.26 & 14.07 & SD \\
NGC 4258 & Sb(s)II & 6.4     &	8.53  & 0.23  & $-$18.91 & 0.36  & H$_2$O Masers \\
Milky Way& (Sc)    & 0.008   &	$-$   & $-$   & $-$17.79 & 0.027 & SD \\
NGC 4342 & E7	   & 16      &	13.37 & 1.0   & $-$17.65 & 3.20  & SD \\
NGC 221  & E2	   & 0.72    &	8.72  & 1.0   & $-$15.57 & 0.03  & SD \\
\hline
\end{tabular}
\end{center}
\end{table}
\end{quote}

The $K$-band luminosity of our Galaxy's bulge, $L_K = 1.2 \times
10^{10}~L_{\odot}$ (Kent et al.\ 1991), was converted to a $B$-band
luminosity by assuming that the galactic bulge has the same color as
M31's bulge at 1 kpc, ($B - K$)$_0 = 4$ (Battener et al.\ 1986).

One outstanding question is whether or not there is a massive black
hole in the center of every nucleated galaxy.  Although the present
census is too biased and incomplete to answer this question, the
presence of MBHs in three local group galaxies make it clear that
massive black holes are more likely to be the rule than the exception.
A more restricted question is whether or not every active galaxy hosts
a MBH.  All eight of the AGNs in Table 3 have large central dark
masses.  As already noted, the active nuclei in NGC 4258 and the Milky
Way are the two strongest cases for MBHs.  These facts lend powerful
support to the supposition that all AGNs are powered by accretion onto
massive black holes.  Ho (1997, 1998) finds that 40$\%$ to 50$\%$ of
all nearby galaxies have active nuclei, suggesting that there is low
to moderate levels of accretion onto a MBH in roughly half of all
nearby galaxies.  Taking into account the strong evolution of galaxy
activity with redshift, we are again led to the conclusion that most,
if not all, galaxies host a MBH.  Richstone (1998) comes to the same
conclusion.

\begin{figure}[t]
\vbox to 7.5cm{\vspace*{-0.7cm} \hspace*{0.5cm}
\psfig{figure=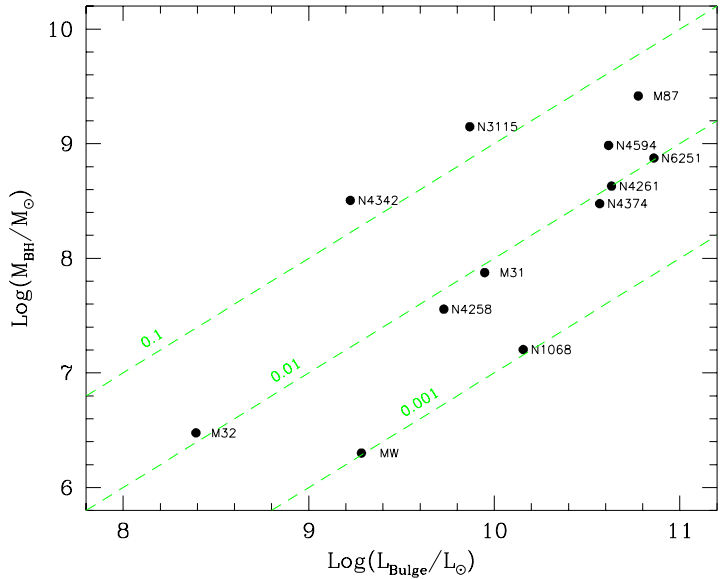,width=11cm}
}
\caption{Black hole mass versus B-band bulge luminosity.  The dashed
lines are the locii of constant ratios of $M_{\rm BH}$/$L_{\rm Bulge}$.}
\end{figure}

Using mostly ground-based data, Kormendy and Richstone (1995) made the
interesting suggestion that there may be a strong relationship between
the mass of a galaxy's central black hole and the luminosity of the
galaxy's bulge.  If true, such a relationship would provide important
clues as to how MBHs form and grow.  To test their suggestion with
more recent data, in Figure 2 we show log($M_{\rm BH}/M_{\odot}$)
versus log($L_{\rm Bulge}/L_{\odot}$). The errors in the masses of the
black holes are not yet well understood.  In most cases we think the
errors are likely larger than the formal errors given by the authors.
However, the masses appear to be good to a factor of two or better.
The luminosities of the bulges in NGC 4258 (Sb(s)II) and NGC 1068
(Sb(rs)II) are derived from Simien and de Vaucouleurs (1986) fraction
of bulge light versus RC3 type, rather than from direct measurements.
However, the bulge/total value of 0.23 is very consistent with Kent's
(1987) value of 0.24 for M31 (SbI-II).

Figure 2 does not support the supposition that there is a strong
relationship between black hole mass and bulge luminosity.  The two
galaxies with the best mass determinations, NGC 4258 and the galaxy,
are separated by nearly a factor of 10 in this diagram.  We further
note that because it is difficult to measure the mass of a ``small''
black hole in a giant elliptical with low central surface brightness,
the lower right hand corner of Figure 2 always will be difficult to
populate.

In the next few years we can anticipate that the number of galaxies
with HST measurements of central masses will increase by factors of
several.  This continuing census should provide definitive answers to
questions about the frequency and ranges of black hole masses in
galaxies.

{\bf Acknowledgments} --- 
This research was supported by NASA grant NAG5-1630 to the Johns
Hopkins University.

\end{document}